\documentclass [11pt]{article}

\usepackage{graphicx}
\usepackage{amssymb}
\usepackage{amsfonts}
\usepackage{multirow}
\usepackage{amsmath} 
\usepackage{longtable}
\date{November 2012, TR-201211213534 (local report no.)}

\begin{document}

\author{
Sonja Kabicher-Fuchs$^1$, Stefanie Rinderle-Ma$^1$, Jan Recker$^2$,\\ Marta Indulska$^3$, Francois Charoy$^4$, Rob Christiaanse$^5$, Reinhold Dunkl$^1$, \\Gregor Grambow$^6$, Jens Kolb$^6$, Henrik Leopold$^7$, and Jan Mendling$^8$ \\
$^1$University of Vienna, Austria \\
sonja.kabicher-fuchs, stefanie.rinderle-ma, reinhold.dunkl@univie.ac.at\\
$^2$ Queensland University of Technology, Australia\\ 
j.recker@qut.edu.au\\
$^3$The University of Queensland, Australia\\
m.indulska@business.uq.edu.au\\
$^4$ Universite de Lorraine, France\\
charoy@loria.fr\\
$^5$Free University Amsterdam, The Netherlands\\
r.christiaanse@efco.solutions.nl\\
$^6$ Ulm University, Germany\\
gregor.grambow,jens.kolb@uni-ulm.de\\
$^7$ HU Berlin, Germany\\
henrik.leopold@wiwi.hu-berlin.de\\
$^8$ Vienna University of Economics and Business\\
jan.mendling@wu.ac.at
}

\title{Human-Centric Process-Aware Information Systems (HC-PAIS)}
\maketitle

\begin{abstract} Process-Aware Information Systems (PAIS) support organizations in managing and automating their processes. A full automation of processes is in particular industries, such as service-oriented markets, not practicable. The integration of humans in PAIS is necessary to manage and perform processes that require human capabilities, judgments and decisions. A challenge of interdisciplinary PAIS research is to provide concepts and solutions that support human integration in PAIS and human orientation of PAIS in a way that provably increase the PAIS users' satisfaction and motivation with working with the Human-Centric Process Aware Information System (HC-PAIS) and consequently influence users' performance of tasks. This work is an initial step of research that aims at providing a definition of Human-Centric Process Aware Information Systems (HC-PAIS) and future research challenges of HC-PAIS. Results of focus group research are presented. 
\end{abstract}

\section{Introduction}
Process-Aware Information Systems (PAIS), such as workflow systems, aim to increase the efficient performance of an organization's processes. However, there seems to be unexploited potentials in research for the development of concepts, implementations, and evaluations that address the inclusion of humans in such systems.

As indicated in \cite{Kabicher-Fuchs:2012}, scattered works address human orientation in PAIS  by considering different aspects such as the interaction of automated and human workflows \cite{agrawal07,ayachitula07}, human interactions \cite{schall07,kim09}, flexibility in workflow enactment \cite{faustmann00}, resource allocation for skill acquisition and diversification in organizations \cite{agarwal11}, business processes understandability \cite{mendling07,melcher10}, and tuning of functionalities in a human-oriented way \cite{vanderfeesten05}. These works contributed more or less towards solutions and research that support and acknowledge a brighter inclusion of humans in PAIS. However, there is far more potential for new innovations than exploited in PAIS research so far. For example, we miss:
\begin{itemize}
\item research in PAIS that deals with motivation and satisfaction of humans working with such systems. 
\item research that provides proven concepts that integrate humans' knowledge, skills, competencies, needs, wishes and goals into PAIS systems. 
\item terminology that puts up an umbrella over all the contributions that aim at including humans into PAIS.  
\end{itemize}

In this paper we conducted focus group research in a preliminary research stage to gather the participants' understanding of Human-Centric Process-Aware Information Systems (HC-PAIS). The goal of this first research step was to define HC-PAIS and to collect future research topics of HC-PAIS. 

\section{Methodology}

In this work focus group research was conducted. Focus groups rely on interaction and discussions within the group directed by a moderator \cite{rea} and which refer to particular topics provided by the researcher \cite{gibbs}. Thereby focus groups aim at drawing on the respondents' attributes, such as attitudes and experiences \cite{gibbs}. Focus groups are organized events \cite{gibbs}. They are focused in the sense that they involve some kind of collective activity \cite{kitzinger}. Focus groups generally involve eight to twelve participants \cite{rea}. They can be used during different phases of the research, such as the preliminary or exploratory stage of a study \cite{gibbs}. Focus groups can be considered as an information-gathering technique and as research tool for in-depth qualitative research \cite{rea}. Advantages of focus groups are, for example, the interaction between participants and the elicitation of (multiple) views and understandings \cite{gibbs}. The benefit of focus groups to participants are, for example, the opportunity to be involved in decision making processes, and to be valued as experts \cite{gibbs}. 

\subsection {HC-PAIS Focus Group (FG)}

The HC-PAIS focus group was conducted in June 2012 during the working session of the 1st International Human-Centric Process-Aware Information System (HC-PAIS) Workshop and the Workshop on Governance, Risk and Compliance (GRCIS) which were held in conjunction at the CAiSE 2012. The working session comprised eight participants and two members of the research team (one of them took the role of the moderator). Participants of the working session were mainly authors and presenters of the HC-PAIS and GRCIS workshop papers. We consider the working session participants as a kind of focus group based on the commonality that each participant published a scientific contribution that considered to some extent humans in the context of Business Process Management or Process-Aware Information Systems (e.g. stakeholder-centric modeling, human-centric workflows, human-centric abstraction, natural language transformation into process models, process model understandability, patients, practitioners, user and developer perspectives) via the publishers ACM, Elsevier, IEEE, and Springer in the period 2006-2012.   

The goals of the focus group were (a) to define what is meant by human orientation in Process-Aware Information Systems, and (b) to collect future research topics in the context of Human-Centric Process Aware Information Systems. To address the first goal (a), focus group participants were asked to elaborate a flipchart in a small group to the question 'What do you understand by human-orientation in PAIS?'. The flipcharts were then presented to the focus group members and were discussed. To address the second goal (b), focus group participants were asked to individually write three research topics of human-orientation in PAIS on moderation cards (one topic per moderation card). The participants should consider a call for papers for a conference that concentrates on human-orientation in PAIS. The moderation cards were then pinned on a board and explained to the focus group by each participant. 

\section{HC-PAIS Definition and Research Topics}

In this chapter we present results of the HC-PAIS focus group. The first step of the focus group towards defining human orientation in PAIS was to brainstorm to the question 'What do you understand by human-orientation in PAIS?' in three small groups. The lists of ideas that were spontaneously contributed by the members are illustrated in Table \ref{tab:brainstorming}.

{\footnotesize
\begin{longtable} {|| p{4.5cm}| p{3.5cm}| p{3cm}||}
Group 1 & Group 2 & Group 3\\ \hline
\endfirsthead
Group 1 & Group 2 & Group 3\\ \hline\hline
\endhead
& & \\ \hline
\endfoot
\hline
\caption{FG - What do you understand by human-orientation in PAIS?}
\label{tab:brainstorming}
\endlastfoot
- supportive, not restrictive & Tasks: & Usability\\
- ease of use & - Commitments & -Efficient\\
- adaptability, flexibility & - Working & - Acceptance\\
- customized to target group/ domain/ culture/ expertise & - Deciding & - Ergonomic\\
- context-aware & - Notifying & - Individualization of processes \\
- be-directional & - Crowdsourcing & - Patterns out of individual processes\\
- granularity & Theory: & \\
- not restricted to business & - cognitive fit/load & \\
- diversity & - institutional economies & \\
- connectivity & - cultural dimension & \\
- motivation & - activity theory & \\
- privacy, fear, control & - expertise & \\
- development & - leadership/ mgmt & \\
- efficiency & - Speech Act Theory & \\
- creativity & - Language Action Perspective & \\
- (explicit) feedback & Roles: & \\
- self-adapting & - process owner & \\
- interface design & - process analyst & \\
- multiple devices and mobility & - process participants & \\
- collaboration & - system engineer & \\
- knowledge dissemination & - knowledge sharing & \\
- error proof & - Human-Centric PAIS & \\
- ad-hoc & - end user modeling & \\
- pockets of 'dynamicity' & - collaboration & \\
- audit trail & - culture & \\
- guidance (adaptive level) & - motivation & \\
- variable prescriptivity & - principle-agent & \\
- communication btw. users & - politics & \\
 & - governance & \\
 & - adaptation & \\
 & - cognitive fit & \\
 & - social aspects & \\
\end{longtable}
}

After the brainstroming session in small groups, the flipcharts were presented to all participants of the working session. The flipcharts were the basis for the next phase in which jointly ingredients of the definition for Human-Centric Process-Aware Information Systems (HC-PAIS) were brought together. The result of the defining phase is illustrated in the following. 

\begin{figure}
	\centering
	\includegraphics [width=0.8\textwidth]{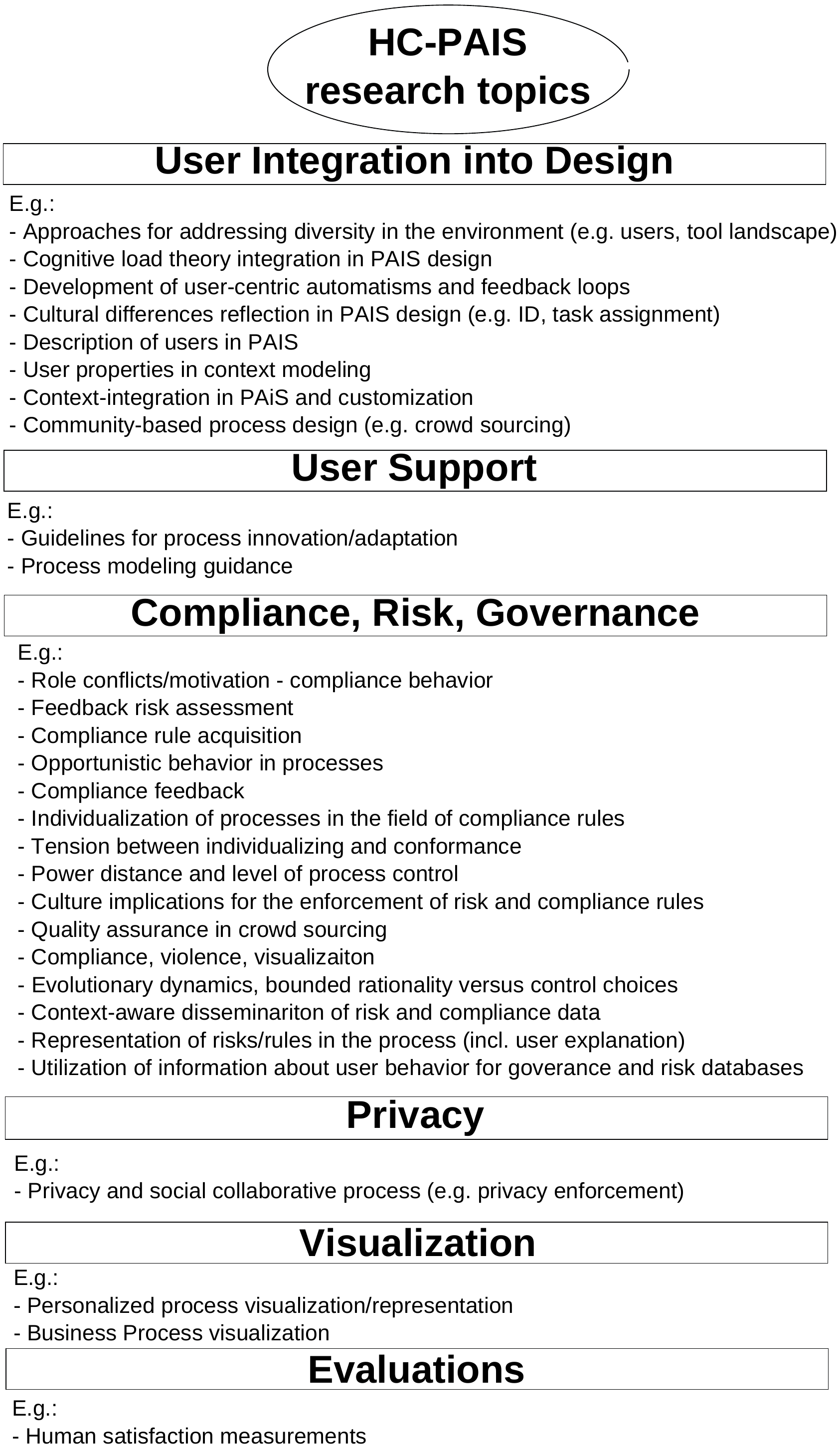}
  \caption {\label{fig:hcpaisResearchTopics} HC-PAIS Research Topic Collection}
\end{figure}

\newtheorem{xyz}{Definition} 
\begin{xyz}[Human-Centric Process-Aware Information System] 
A Human-Centric Process-Aware Information System (HC-PAIS) is rather qualifying than defining an Information System. The consequence of a HC-PAIS will be happiness, motivation, and satisfaction and consequently better performance of humans working with the HC-PAIS. The focus of a HC-PAIS is on process participants. Process participants are humans working with the PAIS to perform organization's processes. The qualification of Information Systems for HC-PAIS is connected to technology acceptance. HC-PAIS is the behavioral perspective on (a) human interaction, (b) human-system interaction, and (c) their interaction in processes, facilitated by PAIS.   
\end{xyz}

In the next phase, ideas of future research topics in the field of Human-Centric PAIS were collected in the focus group. Focus group members were asked to mention three future research topics of Human-Centric PAIS and three future research topics that link the areas human orientation in PAIS and governance, risk and compliance. The pinboard moderation method led to a list of topics. The collected research topics were grouped into different fields, as illustrated in Figure \ref{fig:hcpaisResearchTopics}.

\section{Conclusion}

This contribution is a preliminary step of a research study that concentrates on Human-Centric Process-Aware Information Systems (HC-PAIS). In this work we present results of a focus group that aimed at providing a definition of HC-PAIS and future research challenges of HC-PAIS. Future work will include a Delphi study with selected experts in the field of human resources and resource allocation in PAIS to provide a collection of current and future research challenges of HC-PAIS in a more validated way. Furthermore, we will continue our research \cite{Kabicher-Fuchs:2012} that aims at providing concepts that integrate human capabilities (such as experiences, skills, qualifications) into PAIS in a reliable and contemporary way to use them for task allocation that supports users' development goals.  

\bibliographystyle{plain}
\bibliography{HCPAISRefs}

\end{document}